\documentclass[11pt,a4paper,twocolumn,notitlepage]{article}
\usepackage{amssymb,amsmath}
\usepackage{amsthm}
\usepackage{bm}
\usepackage[french]{babel}
\usepackage[latin1]{inputenc}
\usepackage[geometry,weather,misc,clock]{ifsym}
\usepackage{hyperref}
\hypersetup{pdftex, colorlinks=true, linkcolor=blue, citecolor=blue, filecolor=blue, urlcolor=blue, pdftitle=DWs, pdfauthor=vdolocan , pdfsubject=, pdfkeywords=}
\usepackage[pdftex]{graphicx}

\makeatletter

\begin{document}

\title{Domain wall pinning and interaction in rough cylindrical nanowires}

\date{}
\maketitle

\author{Voicu O. Dolocan\footnote{ \emph{Email:} voicu.dolocan@im2np.fr} \\
Aix-Marseille University, Marseille, France \and IM2NP CNRS, Avenue Escadrille Normandie Niemen, 13397 Marseille, France
}

\begin{abstract}
Interactions between pairs of magnetic domain walls (DW) and pinning by radial constrictions were studied in cylindrical nanowires with surface roughness. It was found that a radial constriction creates a symmetric pinning potential well, with a change of slope when the DW is situated outside the notch. Surface deformation induces an asymmetry in the pinning potential as well as dynamical pinning. The depinning fields of the domain walls were found generally to decrease with increasing surface roughness. A DW pinned at a radial constriction creates a pinning potential well for a free DW in a parallel wire. We determined that trapped bound DW states appear above the depinning threshold and that the surface roughness facilitates the trapped bound DW states in parallel wires.

\noindent \textbf{Pacs}: 75.60.Ch, 75.78.Cd, 75.78.Fg

\end{abstract}

\vspace{0.2in}

Nowadays, tailoring materials at the nanoscale to fabricate devices with precise functionality is an intensive area of research. A multitude of magnetic nanostructures are proposed as storage or logic devices. This devices are mainly based on moving magnetic domains separated by domain walls (DW). The motion of DWs can be controlled by magnetic field or electric current\cite{Allwood,Parkin,Hayashi}. 

Typically, device applications use ferromagnetic nanostrips (flat nanowires). To reliably control the movement of DWs in such structures, the DWs are usually pinned at notches\cite{Klaui} although other geometrical imperfection as kinks\cite{Glathe1}, antinotches\cite{Petit1} or protuberances\cite{Franchin} were investigated. Enhanced DW propagation due to the edge roughness\cite{Nakatani,Martinez1} or internal disorder\cite{Wiele} was equally predicted. The pinning potential was found to depend on the DW type (transverse or vortex)\cite{Petit2}. For a transverse wall, the constrictions create a single potential well. 

Cylindrical nanowires present an alternative to the nanostrips. Transverse walls (TDW) in cylindrical nanowires present some interesting properties as the Walker limit is completely suppressed and theirs velocity and precession speed depend linearly on the applied current\cite{Yan2}. They propagate without deforming theirs internal structure and are mainly studied for microwave oscillators based on DW. Very few studies investigated the pinning and interaction of DWs in such geometry\cite{Franchin,Dolocan3}. Arrays of cylindrical nanowires can be easily fabricated using electrodeposition. One way to create geometrical defects in such wires is by diameter modulation\cite{Pitzschel}, therefore rotationally symmetric constrictions can be fabricated.

In this paper, the interaction between TDWs and the pinning of a TDW by a rotationally symmetric constriction (notch) in ideal and realistic (surface rough) cylindrical ferromagnetic nanowires are studied. Our results establish that a radial notch produce a symmetric pinning potential with a change of slope when the DW is outside the notch. When surface roughness is present, it is found that the pinning potential becomes asymmetric and that the depinning field generally decreases. Furthermore, a pinned DW at a notch produces a pinning potential for a free DW situated in a parallel wire (magnetostatic pinning). Trapped bound states of two DWs in parallel wires can form above the depinning threshold, due to large oscillations in the spatial position. Surface deformation increases the probability of the trapped DW states.  This can strongly influence the control of DWs in cylindrical nanowire based devices. The influence of the temperature is discussed using the escape rate theory and the stochastic 1D model. Our results show the importance of understanding the pinning mechanism and of mutual interaction between DWs in cylindrical nanowires.


\begin{figure}[b!]
  \includegraphics[width=7cm]{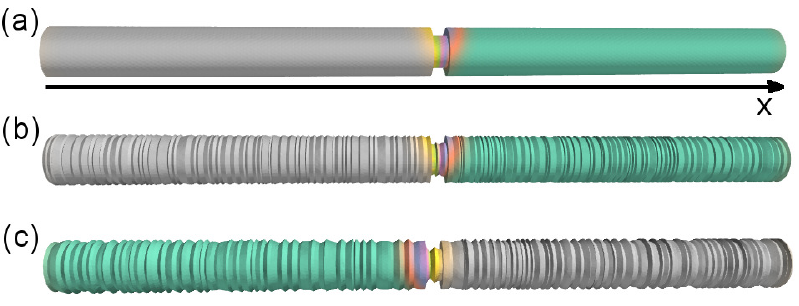}\\
 \caption{\label{Fig.1} (Color online) Cylindrical nanowires with a radial constriction used in simulations: perfect cylinder (a) and cylinder with radial surface roughness with amplitude of 2nm (b) and 4nm (c). The equilibrium position of a pinned DW is shown in each case.}
\end{figure}


To investigate the dynamics of DWs, 3D micromagnetic simulations were computed using the Nmag package\cite{Fischbacher} which determine the spatial distribution of the magnetization dynamics\cite{Dolocan1} solving the Landau-Lifschitz-Gilbert equation. Fig.~\ref{Fig.1} shows the simulated structures. Cylinders with a length of 900nm and 60nm in diameter were used. The cylinders were discretized into a mesh with a cell size of 4nm, inferior to the exchange length ($\sim$5nm for Ni). We consistently checked on a number of test cases that smaller cell size discretization (3nm) does not influence the results presented below. We use the material parameters of Nickel: saturation magnetization $\mu_0$M$_{s}$=0.6T, exchange stiffness A = 1.05 $\times$ 10$^{-11}$J/m, $\gamma$=188.5GHz/T (g factor of 2.15), damping parameter $\alpha$=0.015. No magnetocrystalline anisotropy is considered and the temperature is set to T = 0K. The stable structure in our nanowires is a V-shaped TDW separating domains pointing along the cylinder axis ($x$ axis). The average position of the DW center ($x$) is extracted for each applied field along with the azimuthal angle ($\psi$) of magnetization in the $yz$ plane. 

The constrictions were created as a sharp modulation of the nanowires diameter, having a length of 20nm (same as wall width) and 10nm in height. The surface roughness (Fig.~\ref{Fig.1}(b), (c)) was created from a random profile along the cylinder axis, with a predefined amplitude deformation $\sigma$ and variable correlation length $\lambda$, full revolved around $x$-axis therefore preserving the rotational symmetry. Two radial deformation were used: a soft roughness with $\sigma$=2nm and $\lambda$ between 3nm and 10nm, and a hard roughness with $\sigma$=4nm and $\lambda$ between 3 and 15nm.



\paragraph{Single notch, perfect wire.}\label{singlenotch}


\begin{figure*}[t!]
  \includegraphics[width=13cm]{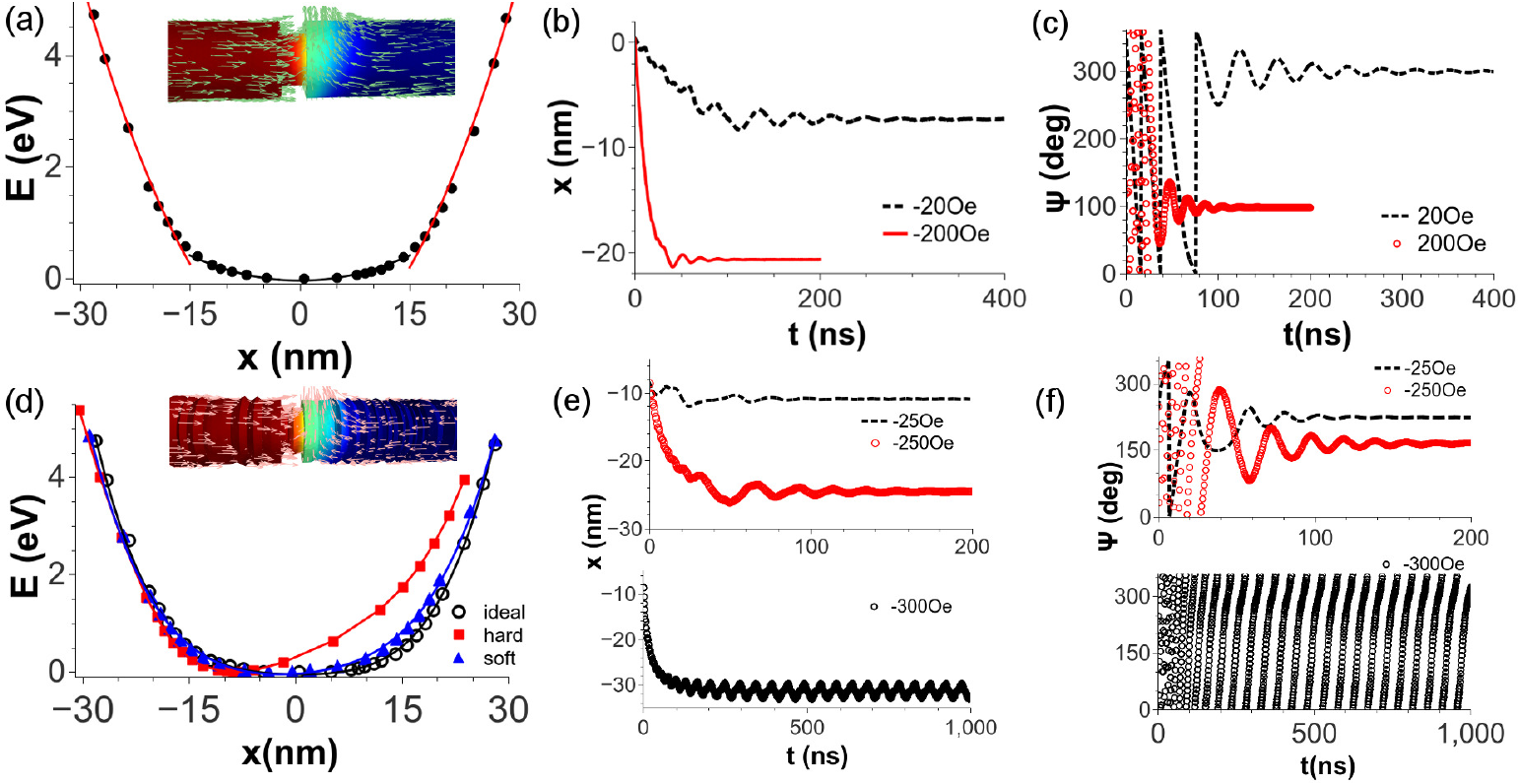}\\
 \caption{\label{Fig.2} (Color online)  Potential pinning energy of a concentric notch situated in an ideal (a) and rough nanowire (d). The insets show the DW position for an applied magnetic field of 200Oe. The lines in (a) represent fits with harmonic potentials, in (d) are guide to the eyes. (b) and (c) show the DW position and DW precession angle for two dc applied fields for the ideal case and (e) and (f) for three dc applied fields for the rough wire.}
\end{figure*}

First, the pinning of a DW by a single regular notch situated in the center of a cylindrical nanowire was investigated. To determine the pinning potential, a dc magnetic field was applied and the distance traveled by the DW (at equilibrium) was computed. The normalized potential energy (exchange + demagnetizing) is represented in Fig.~\ref{Fig.2}(a). The potential profile is symmetric with a change of slope around $x$=10nm, deviating from the simple harmonic potential of symmetric notches in flat nanowires\cite{Martinez2,supp}. When most of the DW is inside the radial notch (lower elongation x$<$10nm), the DW position varies linearly with applied field (see details in Ref.\cite{supp}) and the pinning potential is harmonic with a spring constant $k_{l}=3.21\times 10^{-4}$N/m.  Above $x$=10nm, when most of the DW is outside the notch, the variation of DW position is still linear but with a higher spring constant $k_{h}=1.23\times 10^{-3}$N/m as the DW starts to deform\cite{Yuan}. The maximal equilibrium distance from the center of the notch is 30nm before depinning at H$_{dep}$=$\pm$300 Oe. While it travels to the equilibrium position the DW also precesses around the cylinder axis as shown in Fig.~\ref{Fig.2}(b) and (c). The relaxation time can be as long as 600ns. As the dc field increases, the DW tends quicker to the equilibrium position and the oscillation damps faster.   

It was found that the pinning profile of the radial notch does not depend on the chirality of the TDW. It is identical for head-to-head and tail-to-tail DW pinned at the notch. The profile is also identical if the external perturbation is a dc spin polarized current (polarization P=0.7, non-adiabatic parameter $\xi$=0.05), although the magnetization in the DW precesses in a steady state (no damping of $\psi$ due to the spin torque compensation). In this case, the depinning current is 2.4$\times 10^{10}$ A/m$^2$.
 
The domain wall width suffers a small modification when submitted to the external field (inset of Fig.~\ref{Fig.2}(a)). The width contracts by 2 nm just before depinning from 20nm (without field) to 18nm suggesting a small deformation of the wall when pushed by the external perturbation.

\paragraph{Single notch, rough wire.}

To account for imperfections in the wire, we studied the effect of radial surface roughness on the pinning by a notch of a DW. The difference between the two imperfect wires (as in Fig.~\ref{Fig.1}) and the perfect one is represented in Fig.~\ref{Fig.2}(d). We observe that the soft surface roughness introduce a low asymmetry in the pinning potential of the notch. Without external perturbation, the DW is pinned inside the notch at a local pinning site, which is not in the center of the notch. The elongation from the equilibrium position varies with the local complex potential present. Even the soft roughness already modifies the depinning field which is slightly asymmetric (+270Oe,-280Oe) and smaller than for perfect wires.

When hard roughness is present, the asymmetry in the pinning potential can increase highly and the local pinning sites outside the notch can trap the DW. In Fig.~\ref{Fig.2}(d), a case is shown of hard roughness that presents a large asymmetry. The equilibrium position of the DW without applied field is situated almost at the left edge of the notch. The depinning field is highly asymmetric, and lower in one direction than the perfect case. The asymmetry in the depinning field is of 70Oe (+230Oe,-300Oe) and the relaxation time varies (Fig.~\ref{Fig.2}(e) and (f)). The oscillation of the DW position, at higher elongation, can take longer time to damp as the local potential depends on the local geometry of the roughness. At -300Oe, just before depinning, the DW is resonantly excited (frequency 20MHz) by the applied field as shown in panels (e) and (f). The amplitude of the spatial oscillation is 5nm, while the DW magnetization precesses around $x$ axis. The resonance frequency can be measured experimentally and the local shape of the potential can be determined\cite{Bedau}. When the DW depins, it jumps from a local pinning position to another while the magnetization can change its direction of precession as the DW can move backward. 

Several cases of hard roughness were studied (the most representative are compared in Ref.\cite{supp}). In each of them an asymmetry in respect with the center of the notch was obtained. The asymmetry in the pinning potential was found to depend mostly on the roughness inside the notch which creates a local equilibrium position which is not in the notch center. The asymmetry is more pronounced when the zero field equilibrium position is close to the notch edge. The depinning field was found to decrease, compared to the ideal case, at least for one field direction. Sometimes, when the DW is pinned at an edge of the notch (without applied field) the depinning field increases slightly (10Oe) for one direction. 


\paragraph{Parallel ideal wires.}


\begin{figure*}[t]
  \includegraphics[width=15cm]{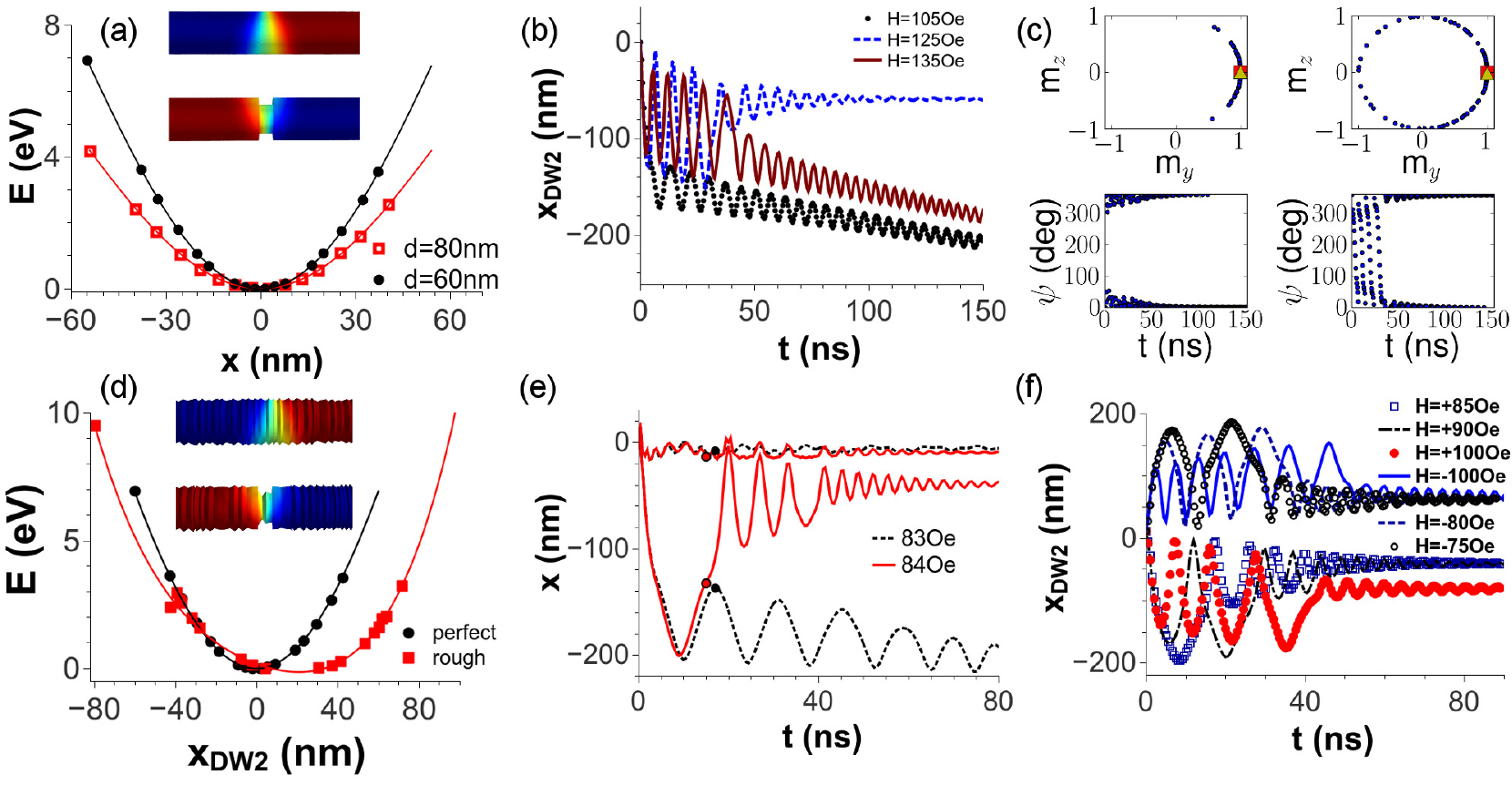}\\
 \caption{\label{Fig.3} (Color online)  (a) Interaction energy, E, between a free TTDW and a pinned HHDW situated in parallel perfect wires as a function of longitudinal displacement. The inset shows the equilibrium position of DWs. (b) Oscillation of the position of upper DW (x$_{DW2}$) for different applied magnetic fields above the depinning threshold. The rotation of the magnetization of the two DWs (in the $yz$ plane) and the azimuthal angle are shown in (c) for H=+125Oe corresponding to the bound state. The squares depict the initial position and the triangles the final position. (d) Comparison of E for smooth and rough wires separated by 60nm. The lines are guides to the eyes. The inset shows the equilibrium position for rough wires. (e) Oscillation of DWs position at and above the depinning field in rough wires. The circles correspond to the instant of first parallel orientation of magnetization in both DW. (f) Oscillation of the upper free DW position in the rough wire for different applied magnetic fields which lead to trapped states.}
\end{figure*}

Next, the interaction between DWs situated in two separated nanowires was investigated: a free tail-to-tail DW (TTDW) situated in a perfect wire and a head-to-head DW (HHDW) pinned at a notch (as above) situated in the center of a lower wire (Fig.~\ref{Fig.3}(a) inset). The wires are separated by a distance $y$=60nm or $y$=80nm. In this configuration, the free DW is trapped in the magnetostatic potential created by the pinned DW. This type of magnetostatic pinning in neighboring nanowires was demonstrated experimentally in nanostrips\cite{Hayward1,Hayward2,OBrien}.

The interaction is detailed in Fig.~\ref{Fig.3}: in (a) the interaction energy is represented function of the horizontal distance $x$ between the two DW when an axial static magnetic field is applied. The potential well is symmetric and the interaction energy between the DWs can be fitted\cite{supp} with the analytical expression of the multipole expansion of Ref.\cite{Kruger}. The free DW in the upper wire has its magnetization pinned in the direction of the magnetization of the pinned DW ($y$ direction) when the magnetic field is inferior to the depinning field (H$_{dep}=\pm$105Oe at $y$=60nm and $\pm$75Oe at $y$=80nm), even if initially its position oscillates with high amplitude (90nm). Above the first depinning field, the spatial oscillations of the free DW (panel (b)) are complemented by full precession of the DW magnetization in the $yz$ plane and the DW is no longer trapped by the pinned DW. However, at +125Oe ($y$=60nm) there is a particular oscillation of the two DWs that finalizes in a bound state, even though the free DW does not have its magnetization pinned initially in the $y$ direction. As can be observed in Fig.~\ref{Fig.3}(c), the free DW magnetization precesses four times before being captured by the lower DW and have its magnetization pinned in the $y$ direction. The particular initial oscillation frequency of the free and pinned DW (see also panel (e)) makes possible such a state in cylindrical nanowires. Above or below 125Oe, the oscillation frequency of the two DW does not lead to a bound state. The energy of the bound state is twice the energy of the two DWs just before depinning (see panel (a)). Any DWs state with energy lower than 7eV and higher than 3.5eV (on the fitted curve) can be a potential bound state, if the initial oscillation (spatial and angular) of the two DWs leads to a parallel orientation of the DWs magnetization at the closest approach. This DW bound state above the depinning threshold does not depend on the separation distance between the wires. If the wires are separated by 80nm, a bound state at 80Oe appears above the the first depinning threshold ($\pm$75Oe).

\paragraph{Parallel rough wires.}

To determine the influence of surface roughness on the DWs interactions, we considered wires with hard surface deformation as before ($\sigma$=4nm, $\lambda$=3-15nm). As expected, even at zero applied field, the DWs relax at local potential minima giving different relaxation positions than the ideal wires. The pinning potential well is wider and symmetric around $x_{DW2}$=15nm, in the case presented in Fig.~\ref{Fig.3}(d), for a separating distance of 60nm. The free DW can travel longer distances during the initial transient oscillation before being captured by the lower DW, and its equilibrium position can be as much as twice the maximum equilibrium position in smooth wires. The free DW is considered trapped by the lower pinned DW, when it stabilizes at an equilibrium position having its azimuthal angle in the same direction ($y$) as the lower DW. The height of the pinning potential well diminishes from 3.5eV to 2.5eV before the first depinning threshold. The depinnig field is asymmetric, H$_{dep}$=-69Oe and +83Oe, and corresponds to the first large oscillation that results in precession as in the smooth case. In panel (e), we detail the oscillations of the pinned and free DW at and above the pinning threshold. The filled and empty dots represent the time stamps where for the first time the magnetization of both DWs are parallel. At 84Oe, the free DW oscillates faster and comes closer (by 20nm) to the pinned DW at their first alignment than at 83Oe and therefore synchronize afterward. We deduce that the potential well length is of 120nm, if the DWs are separated by longer distances when they align the trapped state can no longer form.

The surface deformation produce a notable difference: in smooth wires, only a bound state was obtained above the depinning field; in rough wires, we obtained a large number of bound states above the depinnig threshold (Fig.~\ref{Fig.3}(f)). Before being captured, the free DW does several large amplitude oscillations and precessions, increasing the number of large transient oscillations every 5Oe. The roughness facilitates longer transient oscillations and the loss of the initial trapping (azimuthal angle in $y$ direction) at lower applied fields and therefore the precession of the free DW. But the roughness also facilitates the trapping of the free DW by the pinned DW (panel (f)). A trapped state is obtained even after seven precessions of the free DW for H=-100Oe. This is probably due to the fact that the free DW, in the rough wire, jumps between local potential minima having a lower precession speed (due to the lower applied fields) than in the smooth case. At $\pm$105Oe, the free DW always depins as the equilibrium position would be to far to synchronize its magnetization with the pinned DW.

The free DW jumps sometimes larger distances with little variation of the external applied field: from 11nm (zero field) to 30nm (-5Oe) and from -42.5 to -79.5nm (90Oe to 100Oe), which indicates deeper local potential minima around these positions. The local deformation (local potential landscape) modifies highly the energy of the trapped state, the state obtained at 100Oe (x$_{DW2}$=-79.5nm) is a high energy state (9.5eV) while the state obtained at -100Oe (x$_{DW2}$=71.5nm) is a low energy state (3.2eV).



The influence of the temperature on the DW can be estimated considering that DW depinning is usually a thermally activated process. The effect of the temperature, on the escape of a DW from a single pinning potential, can be modeled using an Arrhenius-N\'{e}el type law\cite{OBrien,Himeno,Lucassen}:

\begin{equation}
\Gamma = \Gamma_0 e^{-\Delta V/k_B T}
\end{equation}

\noindent where $\Delta V$ is the barrier height, and $\Gamma_0$ the attempt frequency at zero temperature. The value of $\Gamma_0$ is estimated in the range 10$^7$-10$^{12}$Hz\cite{OBrien,Himeno}. Considering an experimental time of 1ms as in Ref.\cite{OBrien}, the magnitude of the potential barrier a DW could surmount varies from 0.23 to 0.51eV. For the case of the radial notch in the perfect wire, this corresponds to a change in the depinning field of 7Oe, as the height of the barrier at 0K is 4.75eV. The temperature will have a stronger impact on the depinning field of the rough wires, of 25Oe in one direction as the potential is asymmetric and the potential barrier is weaker. For DWs in parallel perfect wires, the temperature will change the depinning field by less than 10Oe, while in rough wires it will change it by 20-25Oe.

The depinning probability of a DW from a pinning potential can be also estimated with the stochastic 1D model\cite{Martinez}. Using the same parameters as the micromagnetic simulation\cite{supp}, we obtain that at T=0K, the depinning occurs at 299Oe as an abrupt jump, the problem being deterministic. At 300K, a non null probability of depinning exists already at 285Oe (of 3$\%$) and the probability becomes 100$\%$ at 297Oe. The depinning probability varies on a range of 10Oe, similar with the estimates above. As the potential barriers are high compared with the thermal energy, we do not expect a dramatic change in the depinning field at 300K.

Cylindrical constrictions require large depinning field compared to the propagating fields of DWs. The depinning field of the captured free DW by a pinned DW situated in a parallel wire is of the same order as in nanostrips. However, as no Walker limit is present, trapping of the free DW can be realized above the depinning threshold at higher fields. The roughness facilitates the trapping, which can be used in design devices based on cylindrical nanowires.

In conclusion, we have presented a systematic study of the interaction between TDWs in smooth and rough cylindrical nanowires. The static properties of the pinning potential in a radial constriction have been analyzed. The potential well is symmetric and a change of slope is observed when the DW is situated outside the notch. The relaxation times are longer than in nanostrips. We examined the interaction between two DWs placed in parallel wires and observed that bound states can exists above an initial depinning field, due to the large oscillation of the DW positions.

We also investigated the case of more realistic wires, where surface roughness is present. The deformation amplitude can change drastically the form of the potential well and the depinning field which becomes asymmetric. The interaction between DWs becomes more complex due to the asymmetry, dynamical pinning and temperature. The roughness increases the probability to obtain trapped bound states after initial large transient oscillations.


The author wish to thank L. Raymond and is grateful for the support of the NANOMAG platform by FEDER and Ville de Marseille. The computations were performed at the Mesocentre d'Aix-Marseille University.


\end{document}